\documentclass[aps,prl,twocolumn,groupedaddress,showpacs,reprint]{revtex4-1}
\usepackage{graphicx}
\usepackage{amsmath}
\usepackage{mathrsfs}
\usepackage{amsmath}
\usepackage{amsfonts}
\usepackage{amssymb}
\usepackage{epstopdf}
\usepackage{lineno}
\usepackage{hyperref}
\usepackage{verbatim}
\newcommand{\ket}[1]{{\left| {#1} \right\rangle}}

\newcommand{\Yb}{$^{171}$Yb$^+\ $}
\newcommand{\Ba}{$^{138}$Ba$^+\ $}
\newcommand{\ketu}{$ \left| \uparrow \right\rangle \ $}
\newcommand{\ketd}{$ \left| \downarrow \right\rangle \ $}
\newcommand{\ketU}{$ \left| \Uparrow \right\rangle \ $}
\newcommand{\ketD}{$ \left| \Downarrow \right\rangle \ $}

\begin{document}
\title{Multi-Species Trapped Ion Node for Quantum Networking}

\author{I.~V. Inlek}
\email[inlek@umd.edu]{}
\author{C. Crocker}
\author{M. Lichtman}
\author{K. Sosnova}
\author{C. Monroe}
\affiliation{Joint Quantum Institute and Department of Physics,
University of Maryland, College Park, MD 20742, USA}

\begin{abstract} 

Trapped atomic ions are a leading platform for quantum information networks, with long-lived identical qubit memories that can be locally entangled through their Coulomb interaction and remotely entangled through photonic channels.  However, performing both local and remote operations in a single node of a quantum network requires extreme isolation between spectator qubit memories and qubits associated with the photonic interface.  We achieve this isolation and demonstrate the ingredients of a scalable ion trap network node by co-trapping \Yb and \Ba qubits, entangling the mixed species qubit pair through their collective motion, and entangling the \Ba qubits with emitted visible photons.

\end{abstract}
\date{\today}
\maketitle

\begin{figure}
	\includegraphics[width = 3.3 in]{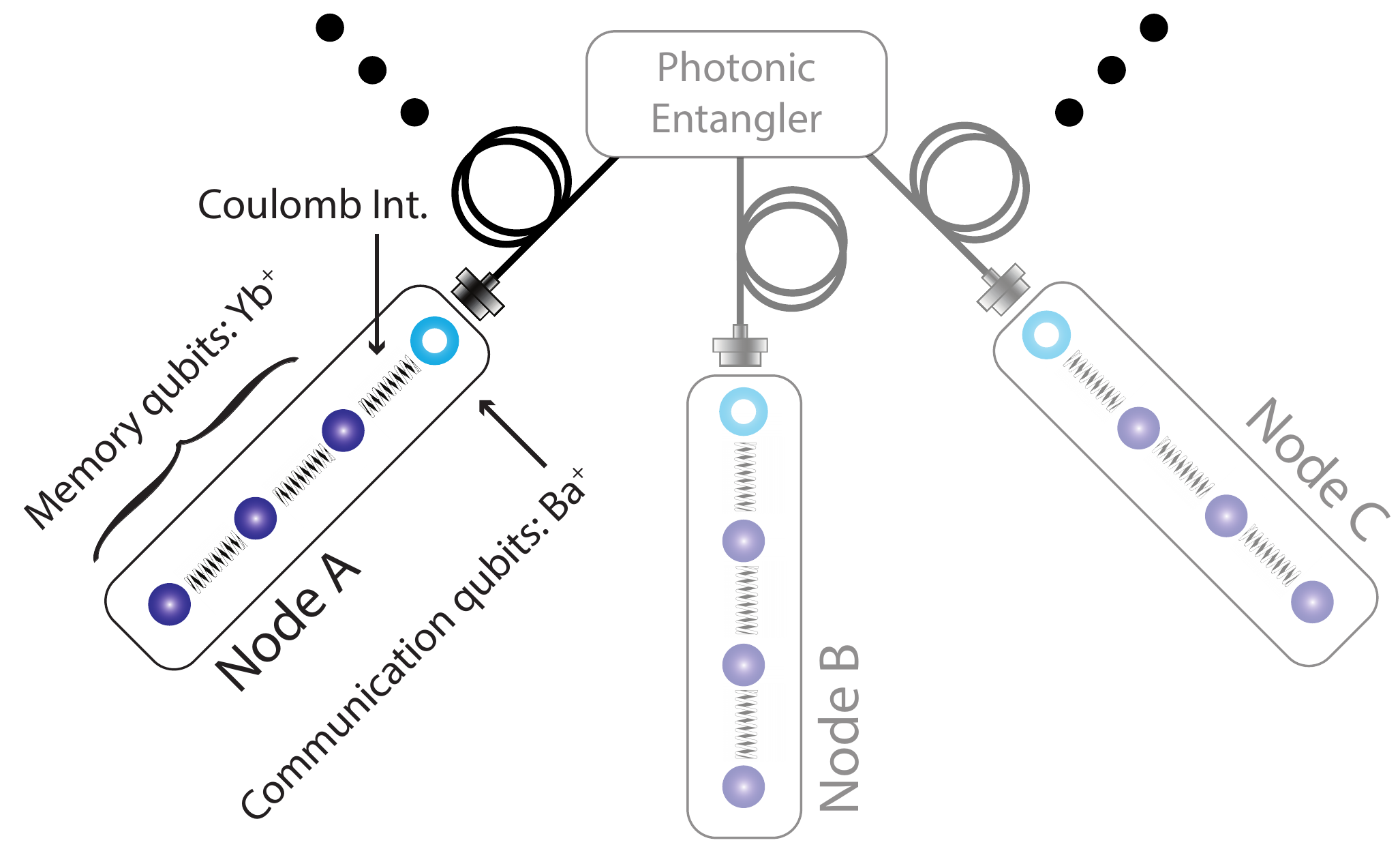}
	\caption{(color online) In an example multi-species ion trap network, \Ba communication qubits are coupled to optical fibers. Using photons entangled with their parent atoms, any pair of \Ba qubits in different nodes can be entangled through a reconfigurable photonic entangler \cite{moehring,musiqc,modular}. Local Coulomb interactions mediate transfer of this entanglement to nearby \Yb memory qubits \cite{QLS,nist-mixed} as well as quantum logic gates within the node \cite{gates_computer}. The disparate electronic transition frequencies of the two species provides the necessary isolation to protect \Yb memory qubits from resonant processes in the \Ba photonic interface.}
	\label{multi_species_module}
\end{figure}

Trapped atomic ions are among the most advanced platforms for quantum information networks, hosting qubit memories that are inherently identical and have unrivaled coherence properties. A single node of the network can be realized with a chain of trapped ions, where local entangling gate operations use external control fields that couple the qubit states through their collective motion \cite{cz,ms,blatt}. Edges of the network can then be implemented by photonic entangling operations between select ``communication" qubits in separate nodes \cite{quantum_internet,quantum_networks_with_trapped_ions, topological_network}. However, the photonic interface for the communication qubits must not disturb the spectator memory qubits, as even a single resonant photon can destroy the quantum memory.  Such isolation is best accomplished by using two different species of atomic ions \cite{normal-modes}; one for local processing and memory, the other for communicating with other nodes as shown in Fig.~\ref{multi_species_module}. 

Here, we demonstrate each of the ingredients of a multi-species ion trap node for use in a potential quantum network \cite{quantum_networks_with_trapped_ions,musiqc}. This includes coherent quantum state mapping between memory and communication qubits, and the generation of photonic qubits entangled with the communication qubits.  We encode the memory qubits in the $^{2}S_{1/2}$ ground state hyperfine ``clock" levels of \Yb atomic ions, $\ket{F=0,m_{F}=0}\equiv\ket{\Downarrow}$ and $\ket{F=1,m_{F}=0}\equiv\ket{\Uparrow}$ \cite{ybqubit}.  As communication qubits, we use the $^{2}S_{1/2}$ ground state electron spin levels of \Ba atomic ions, $\ket{J=1/2,m_{J}=-1/2}\equiv\ket{\downarrow}$ and $\ket{J=1/2,m_{J}=+1/2}\equiv\ket{\uparrow}$ \cite{baqubit}. The \Ba system features relatively long wavelength photon emission lines (493 nm and 650 nm), easing the technological requirements for the photonic interfaces and provides the necessary isolation from the \Yb resonance at 369 nm. We verify the isolation between these two species by observing that the measured coherence time of \Yb qubits ($\sim$ 1.5 s) is not affected by fluorescence or the driving laser light associated with a continuously Doppler cooled \Ba qubit a few microns away. With the application of dynamical decoupling pulses, a \Yb hyperfine qubit coherence time exceeding 10 minutes has been reported in a similar setup where a nearby \Ba ion is used for sympathetic cooling \cite{coherence_mixed}.

\begin{figure}
	\includegraphics[width = 3.3 in]{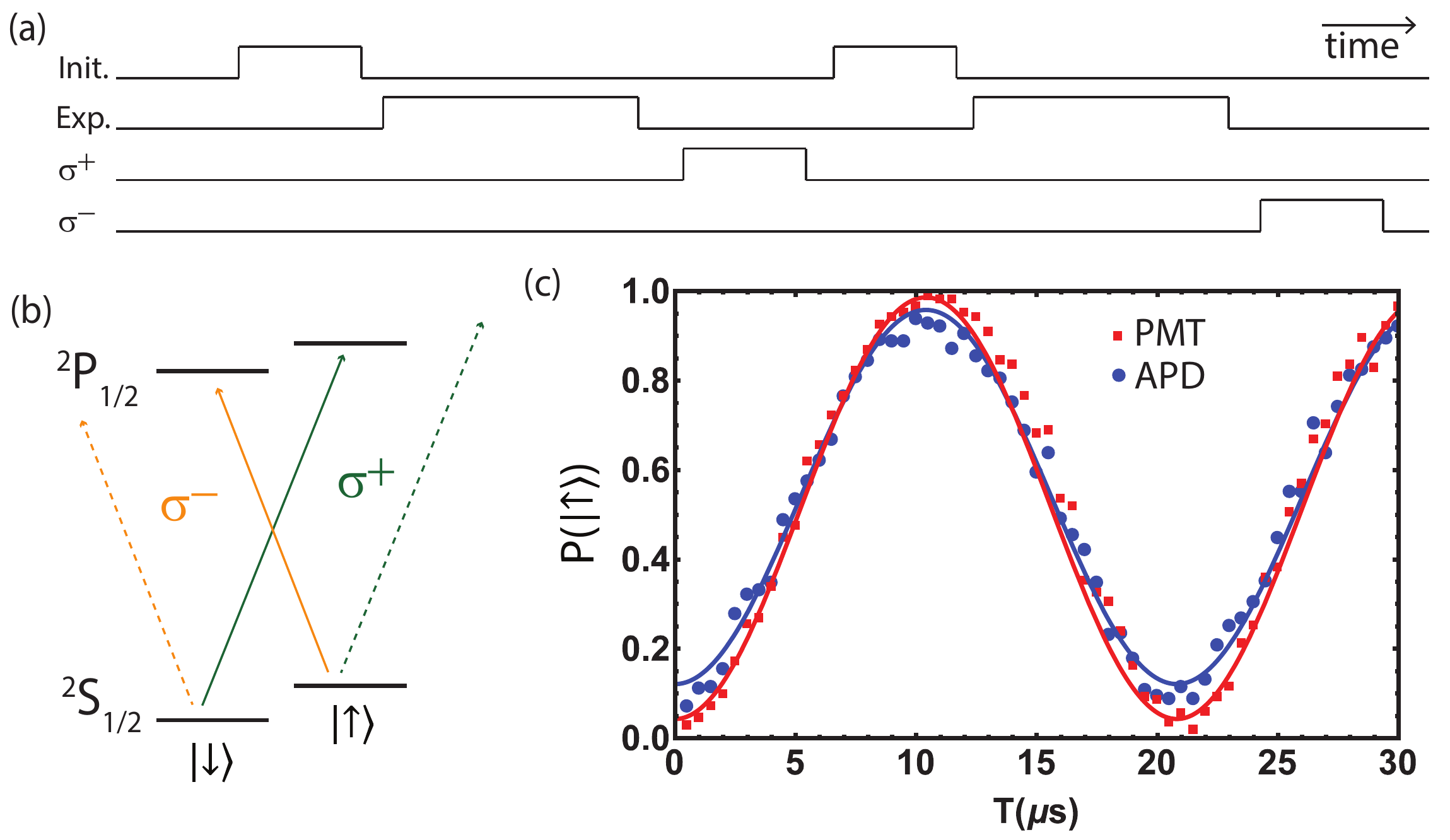}
	\caption{(color online) (a) \Ba detection timing sequence. After initializing to \ketd and preparing the \Ba qubit (an ``experiment"), we excite the $^{2}S_{1/2}$ to $^{2}P_{1/2}$ transition with 493 nm $\sigma^{+}$ polarized light. If the ion is in the \ketd state, on average it scatters $3$ photons before getting optically pumped to the \ketu state, and each photon is detected with a probability of approximately $8\%$.  In contrast, the \ketu state ideally does not scatter any photons due to selection rules. We then repeat the same initialization/experiment steps, but now send $\sigma^{-}$ polarized light, and the situation is reversed with only the \ketu state scattering photons. We cycle between $\sigma^{+}$ and $\sigma^{-}$ detection on identical experiments hundreds of times, and accumulate the collected photons for each detection polarization condition. This allows us to infer the \Ba qubit state without statistical bias or drifts. (b) Relevant energy levels in the \Ba system coupled with $\sigma^{+}$ and $\sigma^{-}$ resonance excitation light at 493 nm. (c) Observed Rabi oscillations between the \ketd and \ketu states, where the experiment consists of stimulated optical Raman transtions between the states with variable duration $T$.  The probability of finding the \Ba qubit in \ketu is extracted from the ratio of the photons collected in the $\sigma^-$ cycle to the total of 300 photons collected with either polarization. The total integration time for the entire $1.5$ Rabi cycles is approximately 2 minutes.  While the effective \Ba qubit detection efficiency is only $8\%$, we estimate a detection accuracy of about $94\%$.}
	\label{detection}
\end{figure}

We use standard spin-dependent fluorescence collection for the near-perfect single-shot detection of the \Yb qubit state \cite{ybqubit}.  The \Ba qubit lacks such an isolated cycling transition, so we detect the \Ba qubit only through averaging many identical experiments.  However, this limitation is not debilitating: in the multispecies network architecture, \Ba qubits serve only as a link between \Yb memory qubits.  Once the \Ba qubit is mapped to neighboring \Yb memories through Coulomb-based gates, quantum information processing does not rely on state detection of the \Ba communication qubits. 
Nevertheless, the fast unbiased detection of \Ba qubits is still useful for calibration and diagnotics of the \Ba system.  We therefore collect \Ba fluorescence with a highly efficient optical objective (NA=0.6, or about 10\% collection) and use an avalanche photodiode with 80\% detection efficiency.   This amounts to an effective $8\%$ single-shot detection efficiency of the \Ba qubit.  
With such low detection efficiency, it is important to minimize sensitivity to slow drifts in photon collection efficiency or the excitation laser.  We therefore collect fluorescence after hundreds of repeated runs of the same experiment by alternating the sense of circular polarization of the excitation laser, as shown in Fig. \ref{detection}.  In this way, we detect both \Ba qubit states independently, providing normalization for the extracted population probability.  Moreover, this technique suppresses statistical bias in the photon collection distribution between the \ketd and \ketu states.

In addition to their use as photonic communication qubits, \Ba ions can be employed in sympathetic cooling of \Yb qubits to maintain occupation in low motional phonon eigenstates for higher fidelity quantum operations. We implement an electromagnetically-induced-transparency (EIT) cooling technique using 493 nm laser beams that are tuned to about 120 MHz blue of the $^2S_{1/2} - ^2P_{1/2}$ transition. These beams introduce a narrow atom-laser dressed state resonance where the red sideband transitions are selectively excited, while blue sideband and carrier transitions are suppressed \cite{EIT-nist,EIT-theory}. With this technique, we cool the out-of-phase motion of a \Ba and \Yb two-ion crystal to $\bar{n}\approx0.06$ and in-phase mode of $\bar{n}\approx0.1$.  

Communication qubits do not require long coherence times as the information can be quickly transferred to memory qubits where it can be stored and used later. However, the short coherence time of Zeeman \Ba qubits, due to high magnetic field sensitivity of about 2.8 kHz/mG, might result in errors during transfer operation. We use an arbitrary waveform generator to apply magnetic field at 60 Hz and higher harmonics with full phase and amplitude control to partially cancel the background field. This technique increases the \Ba coherence time from 100 $\mu$s to approximately 4 ms, which is much longer than any transfer operation gate times.


\begin{figure}
	\includegraphics[width = 3.3 in]{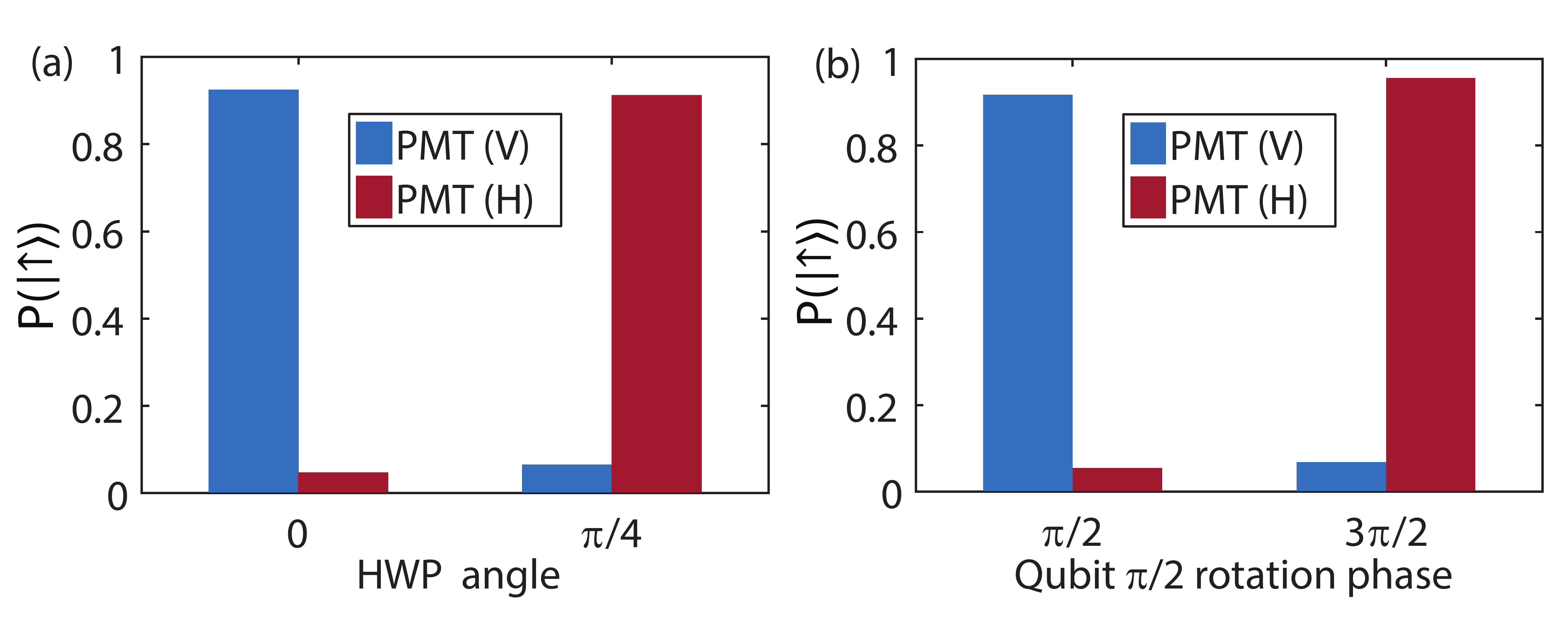}
	\caption{(color online) Correlations between \Ba qubit and emitted photon polarization in multiple bases.
		(a) Measured probability of finding \Ba qubit in \ketu  conditioned upon detecting photon qubit states $\ket{V}$ (blue) or $\ket{H}$ (red). A half wave plate (HWP) rotates the photonic qubit and the data show two measurements corresponding to HWP angles of 0 and $\pi/4$. 
		(b) The photon polarization is rotated by fixing the HWP at $\pi/8$ so that $\ket{H}\rightarrow \ket{H}-\ket{V}$ and $\ket{V}\rightarrow \ket{H}+\ket{V}$. Subsequent photon detection projects the atom to a superposition $(\ket{\uparrow}+\ket{\downarrow})\ket{H} + (\ket{\uparrow}-\ket{\downarrow})\ket{V}$. Following detection of a $\ket{V}$ or $\ket{H}$ photon, we coherently rotate atomic superposition states to $\ket{\downarrow}$ and $\ket{\uparrow}$ with a $\pi/2$ rotation having phase of either $\pi/2$ or $3\pi/2$, recovering high correlations between the qubit and the photon.}
	\label{ion-photon}
\end{figure}

We demonstrate a photonic interface by entangling the \Ba qubit with an emitted photon, through a post-selection procedure 
\cite{ion-photon-cd, ion-photon-ba}.  Here, we initialize the qubit to the \ketd state and weakly excite to the $^{2}P_{1/2}$ $\ket{J=1/2,m_{J}=+1/2}$ level with probability $\textrm{P}_\textrm{exc}\approx 10\%$. The atom decays back to the \ketd state emitting a $\sigma^{+}$-polarized photon, or to the \ketu state emitting a $\pi$-polarized photon. We collect the photons perpendicular to the quantization axis; therefore $\pi$ photons are registered as vertically polarized in this basis ($\ket{V}$) while $\sigma^{+}$ photons are registered as horizontally polarized ($\ket{H}$).  Given that a photon is collected, this ideally results in an entangled state between the \Ba qubit and the photon polarization qubit, \ketd $\ket{H}$+\ketu $\ket{V}$. 

Figs. \ref{ion-photon}(a) and (b) show correlation measurements between atom and photon qubit states in multiple bases, and from these measurements we infer the (post-selected) entanglement fidelity to be $\mathcal{F}\geq 0.86$. We attribute the errors to polarization mixing due to the large solid angle (10\%) \cite{thesis_steve}, multiple photon scattering in the excitation step ($\textrm{P}_\textrm{exc}/4=2.5\%$) and imperfect state initialization/detection (1\%). These error sources can be significantly reduced by collecting photons along the quantization axis \cite{modular,piphoton} and using pulsed lasers for fast excitation of the atom \cite{moehring}.

Next, we demonstrate a deterministic quantum gate between the two species in the node.  We drive coherent Raman transitions in both atomic ions using a single laser for the coherent exchange of information between the \Yb and \Ba qubits.  We show both a direct Cirac-Zoller (CZ) mapping process by resonantly coupling to the collective motion of the trapped ions \cite{cz, QLS} and a dispersive M\o lmer-S\o rensen (MS) quantum gate between the qubits \cite{ms,nist-mixed}.

\begin{figure}
	\includegraphics[width = 3.3 in]{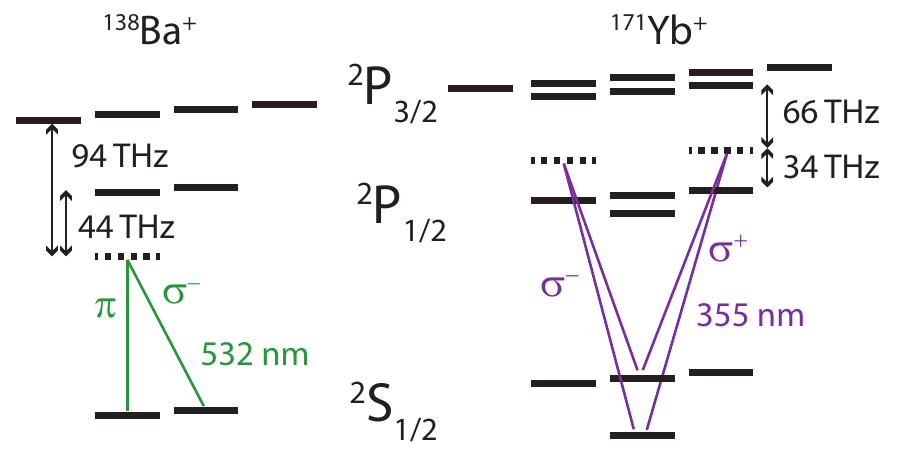}
	\caption{(color online) Off-resonant couplings of 532 nm and 355 nm pulsed laser beams to $^{2}P_{1/2}$ and $^{2}P_{3/2}$ levels in both \Ba and \Yb atomic systems to drive stimulated Raman transitions, with polarizations indicated. Splittings are not to scale.}
	\label{ramans}
\end{figure}

We use a Nd:YVO4 mode-locked pulsed laser (Spectra-Physics Vanguard) to introduce non-copropagating Raman beams that can drive transitions between different vibrational eigenmodes and qubit states \cite{quantum_dynamics}. As shown in Fig. \ref{ramans}, these beams off-resonantly couple to excited levels: the frequency tripled 355 nm output is used for the \Yb system, while the frequency doubled 532 nm output from the same laser is used for the \Ba system. Stimulated Raman transitions are driven when the beat-note frequency between two beams is near the qubit splitting. We choose linear polarizations that are all perpendicular to the quantization axis, allowing desired Raman transitions to be driven while minimizing differential AC Stark shifts on each species \cite{stark_aaron}.  The large bandwidth of the frequency comb easily spans the \Yb qubit frequency of 12.642821 GHz for Raman rotations \cite{comb}. In order to stabilize the beat-note frequency of two Raman beams, we use a feed-forward technique that modulates one of the 355 nm beams to compensate for any changes on laser repetition rate \cite{phaselock}. Since the \Ba qubit splitting is only a few MHz, we do not rely on multiple comb teeth separation for driving transitions on this qubit, hence beat-note stabilization is not necessary on 532 nm beams.  

\begin{figure}
	\includegraphics[width = 3.3 in]{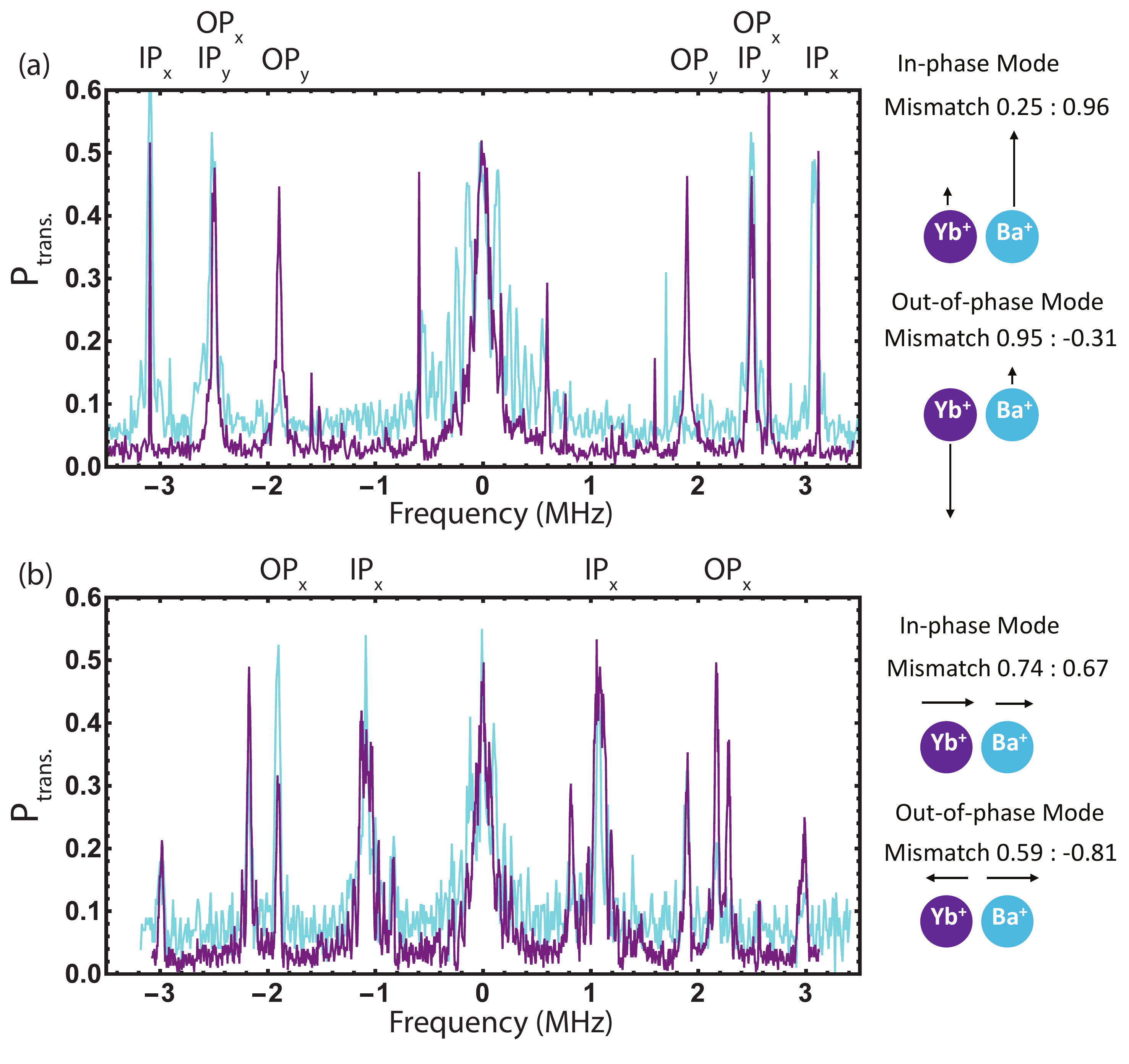}
	\caption{(color online) Raman sideband vibrational spectrum of a co-trapped \Ba-\Yb crystal is shown for (a) transverse and (b) axial directions of motion. The measured probability of changing the qubit state is plotted in blue for \Ba and purple for \Yb, as a function of detuning from the carrier transition where the shared motional phonon state is preserved. The peaks on the positive (negative) values correspond to blue (red) sideband transition in which the spin flip is accompanied with addition (subtraction) of a phonon \cite{quantum_dynamics}.  The sidebands corresponding to the in-phase (IP) and out-of-phase (OP) are labelled for the  transverse (x,y) and axial (z) directions of motion, with their theoretical eigenvector amplitudes indicated at the right.}
	\label{spectrum}
\end{figure}

While the 355 nm (532 nm) radiation nominally couples only to the \Yb (\Ba) qubit, there is a small amount of crosstalk coupling to the other atomic system.  For equal intensities and without regard to the comb spectrum or the light polarization, the \Yb system would feel an effective Rabi frequency from the 532 nm radiation that is $\sim 2.6\%$ of the nominal 355 nm radiation Rabi frequency.  Likewise, the \Ba system would feel a $\sim 11\%$ Rabi frequency from the 355 nm radiation.
However, the required laser polarization and frequency comb spectrum are different for the two atomic qubit transitions, and we exploit this to reduce crosstalk to much less than $1\%$ between the two systems.  The spontaneous Raman scattering rate per qubit Rabi cycle is less than $10^{-5}$ for both atomic species \cite{spontaneous_wes}.  However rare, spontaneous scattering in the \Ba system from 532 nm appears to optically pump the \Ba system through the $^{2}P_{3/2}$ level to the metastable $^{2}D_{5/2}$ state, which has a lifetime of $32$ s \cite{d5/2}.  We overcome these rare pumping events by illuminating the ions with a diffuse 1W orange LED (centered around 617 nm) that excites the $^{2}D_{5/2}$ to $^{2}P_{3/2}$ transition at 614 nm with enough intensity to return the ion to the ground state in approximately 30 ms.

Despite their similar atomic masses, the transverse motion of a coupled pair of \Ba and \Yb ions exhibits a large mismatch in their amplitude for a given mode \cite{normal-modes}, resulting in a smaller motional coupling between the ions (Fig. \ref{spectrum}a).  For this reason, we instead use the better-matched axial modes (Fig. \ref{spectrum}b).  We note that, as the number of ions in the crystal chain increases, the motional eigenvector mismatch in the transverse modes become less significant and these modes can be used conveniently to benefit from higher mode frequencies \cite{transverse_modes}.


\begin{figure}
	\includegraphics[width = 3.3 in]{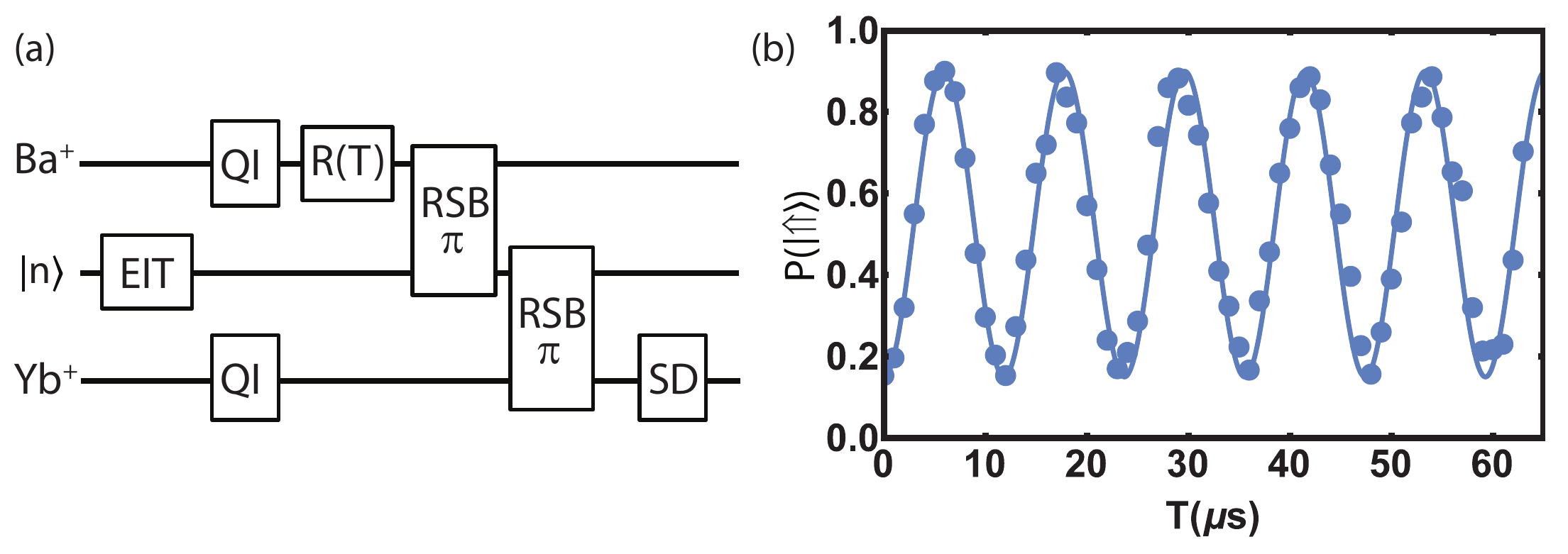}
	\caption{(color online) (a) Experimental steps on mapping the state of \Ba to \Yb using collective motion directly. The procedure starts with EIT cooling, followed by initialization of qubit states (QI) to \ketd and \ketD. Afterwards, a stimulated Raman rotation $R(T)$ of the \Ba qubit over time $T$ prepares the state to be transferred. A red-sideband $\pi$ rotation (RSB $\pi$) on the \Ba qubit transfers this information to shared phonon mode which is then transferred to the \Yb qubit with another red-sideband $\pi$ rotation. In the final step, the \Yb qubit state is measured (SD). (b) The data shows the probability of finding \Yb qubit in \ketU as a function of \Ba qubit rotation time $T$, with an observed state transfer efficiency of $\approx 0.75$.}
	\label{qls}
\end{figure}

We first transfer the qubit state of \Ba to \Yb by directly using the collective motion in a CZ scheme \cite{cz,QLS}. The procedure (Fig. \ref{qls}a) starts with EIT cooling and preparing the state to be transferred onto the \Ba qubit. Next, a red-sideband $\pi$ rotation on the \Ba system transfers information to a shared phonon mode which is then transferred to the \Yb qubit with a second red-sideband $\pi$ rotation on the \Yb system. The overall state transfer efficiency of 0.75, as shown in Fig. \ref{qls}b, was limited primarily by the purity of the initial motional state.  But the main drawback to the CZ method is the necessity of phase coherence between the communication qubit and the CZ mapping operations.  Because the communication qubit may have prior entanglement through the photonic channel, the CZ mapping method requires stabilizing the beam paths to much better than an optical wavelength.

We next demonstrate a M\o lmer-S\o rensen transfer method that relaxes the above limitations.  In the MS transfer scheme \cite{nist-mixed,oxf-mixed}, entanglement and state transfer fidelity only require confinement to the Lamb-Dicke limit \cite{quantum_dynamics}.  Moreover, optical phase sensitivity can be canceled with extra single qubit rotations \cite{nist-mixed} or special beam geometries \cite{plee,phase-insensitive} to achieve phase coherence between photonic entanglement operations and local transfer gates.

\begin{figure}
	\includegraphics[width = 3.3 in]{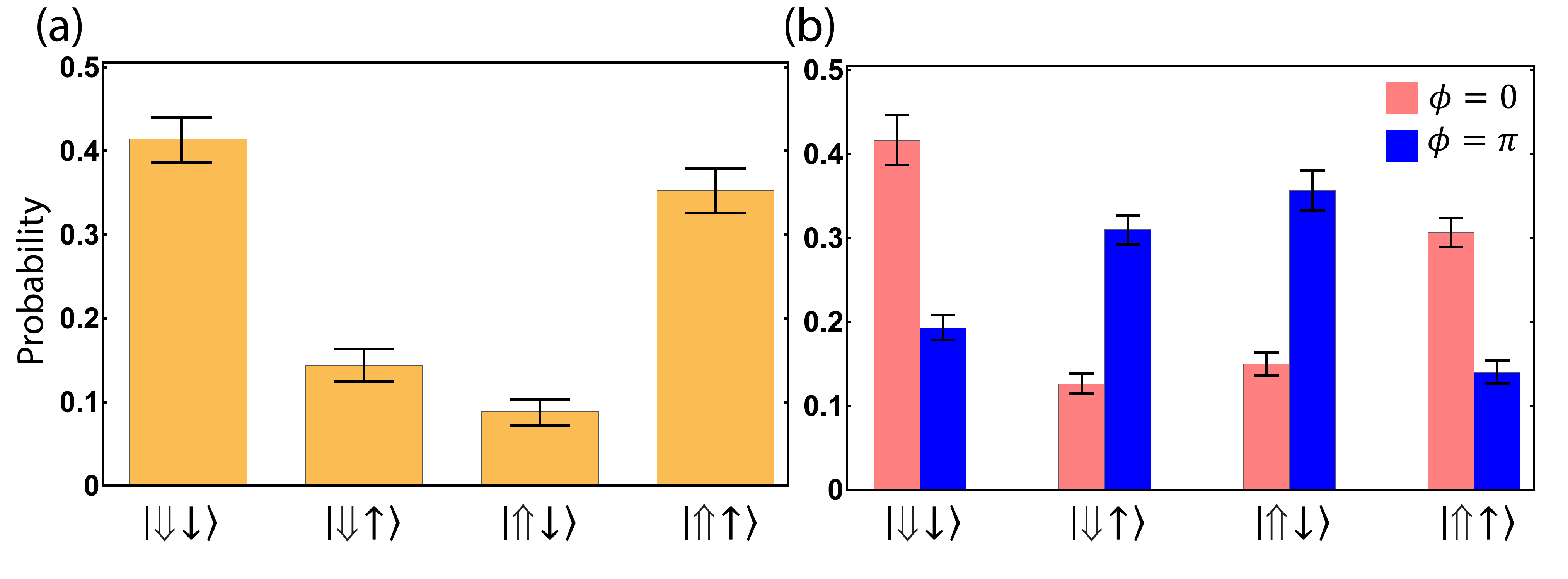}
	\caption{(color online) (a) Measured probabilities of the \Yb and \Ba qubit states after an entangling MS gate is applied to the initial $\ket{\Downarrow \downarrow}$ state. This interaction would ideally create the maximally entangled $\frac{1}{\sqrt{2}} (\ket{\Downarrow \downarrow} -e^{-i \phi_S} \ket{\Uparrow \uparrow})$ state. The optical phases of the driving fields are imprinted on spins with the gate phase, $\phi_S$. (b) Following the MS interaction, $\pi/2$ rotations are applied to both qubits. Phase of \Ba $\pi/2$ pulse is kept constant, while \Yb $\pi/2$ rotation phase is scanned. The data show measured qubit state probabilities at the maximum parity points $|$P($\ket{\Downarrow\downarrow}$) + P($\ket{\Uparrow\uparrow}$) - P($\ket{\Downarrow\uparrow}$) - P($\ket{\Uparrow\downarrow}$)$|$.}
	\label{ms}
\end{figure}

A MS entangling gate is realized in our system by simultaneously addressing sidebands off-resonantly with a symmetric detuning $\delta$ using pairs of non-copropagating Raman beams. Since the pulse pairs of 355 nm and 532 nm follow different paths, they are not necessarily incident on the atoms at the same time. Importantly, a temporal overlap between these pairs is not necessary for the MS interaction; spin dependent forces using the Raman beams can be applied at different times to each atom. The outcome is just a static phase on the entangled state which can be controlled by adjusting either the optical path lengths or the difference between rf beat-note phases of the 355 nm and 532 nm driving fields. These spin dependent forces displace the motional wave-packets of certain two-qubit states in phase space. After a gate time $T=2\pi / \delta$, the motion returns to its original state, picking up a geometrical phase as in the usual MS gate \cite{ms-heating}. We find the correct optical force phase by monitoring the acquired geometrical phases. To maintain shot-to-shot relative optical force phase, we use the same arbitrary waveform generator to drive acousto-optic modulators for 355 nm and 532 nm beams. The fidelity of this operation is approximately $\mathcal{F}=0.60$ as shown in Fig. \ref{ms}, and we attribute this low fidelity to excessive heating ($\dot{\bar{n}}\approx5$ ms$^{-1}$) in the trap \cite{ms-heating}. With higher gate fidelities, two consecutive MS gates with a relative $\pi$ phase difference can be used to reliably swap the qubit states.

Based on the tools demonstrated in this paper, we can extend a quantum network to many nodes using photonic Bell state analyzers to make the photonic connections.  We expect that considerable improvements on the atom-photon and atom-atom entanglement fidelities and rates are possible in order to scale to many interconnected nodes.  First, the encoding of photonic qubits into two different frequencies rather than polarization is expected to provide significant improvements in the remote communication qubit fidelity \cite{teleportation}.  Second, the use of fabricated chip traps with integrated optical elements \cite{integrated_surface_trap} is expected to enhance the connection rate between nodes, but also such traps generally feature better heating rates and positional stability of the trapped ions, allowing for much higher motional gates between memory and communication qubits.

We acknowledge useful discussions with D. Hucul and G. Vittorini.  This work was supported by the ARO with funds from the IARPA MQCO and LogiQ Programs, the AFOSR MURI on Quantum Measurement and Verification, the AFOSR MURI on Quantum Transducers, the ARL Center for Distributed Quantum Information, and the National Science Foundation Physics Frontier Center at JQI.

\bibliography{paperrefs}

\end{document}